\documentclass[%
 reprint,
superscriptaddress,
 amsmath,amssymb,
 prl,
]{revtex4-1}
\usepackage{textcomp}
\usepackage{setspace}
\usepackage{tipa}
\usepackage{verbatim} 
\usepackage{amsmath}
\usepackage{bbold}
\usepackage{subfigure}
\usepackage{amssymb}
\usepackage{accents}
\usepackage{amscd}
\usepackage{graphicx}
\usepackage{epstopdf}
\usepackage{caption}
\usepackage{subfig}
\usepackage{natbib}
\usepackage{color}
\usepackage{braket}

\def\aa{\hat{a}}
\def\ad{\hat{a}^{\dag}}
\def\ads{\hat{a}^{\dag\, 2}}

\def\AA{\hat{A}}
\def\AD{\hat{A}^{\dag}}

\def\be{\begin{equation}}
\def\ee{\end{equation}}
\def\bea{\begin{eqnarray}}
\def\eea{\end{eqnarray}}
\def\nn{\nonumber}
\def\eta{{\it et. al.}}

\definecolor{darkgreen}{rgb}{0,0.4,0}

\begin{document}

\title{Driven Quantum Walks}

\author{Craig S. Hamilton} \email{hamilcra@fjfi.cvut.cz}
\affiliation{%
 FNSPE, Czech Technical University in Prague, B\v{r}ehov\'a 7, 115 19, Praha 1, Czech Republic\\
}%
\author{Regina Kruse}
\affiliation{Applied Physics, University of Paderborn, Warburger Stra{\ss}e 100, D-33098 Paderborn, Germany}
\author{Linda Sansoni}
\affiliation{Applied Physics, University of Paderborn, Warburger Stra{\ss}e 100, D-33098 Paderborn, Germany}
\author{Christine Silberhorn}
\affiliation{Applied Physics, University of Paderborn, Warburger Stra{\ss}e 100, D-33098 Paderborn, Germany}
\author{Igor Jex }
\affiliation{%
 FNSPE, Czech Technical University in Prague, B\v{r}ehov\'a 7, 115 19, Praha 1, Czech Republic\\
}%

\begin{abstract}

In this letter we introduce the concept of a driven quantum walk. This work is motivated by recent theoretical and experimental progress that combines
quantum walks and parametric down-conversion, leading to fundamentally different phenomena. We compare these striking differences by relating the driven quantum walks to the original quantum walk.  Next, we illustrate typical dynamics of such systems and show these walks can be controlled by various pump configurations and phase matchings. Finally,
we end by proposing an application of this process based on a quantum search algorithm that performs faster than a classical search.

\begin{description}
%\begin{comment}
\item[PACS numbers] 05.45.Xt, 42.50.Gy, 03.67.Ac
%May be entered using the \verb+\pacs{#1}+ command. 
%optical couplers 42.82.Et
%Quantum information 03.67.Ac 
%Random walks 05.40.Fb
%coupled oscillators 05.45.Xt
%phase coherence quantum optics 42.50.Gy
%\end{comment}
\end{description}

\end{abstract}

\maketitle

Quantum walks (QW) have become widely studied theoretically  \cite{VenegasAndraca:2012p8540, Reitzner:2011p9515, Kempe:2003p1645} and experimentally in a variety of different settings such as classical optics \cite{Schreiber:2010p5461, Hamilton:2011p4018, Schreiber:2012p5592, Grafe:2012p7498}, photons in waveguide arrays \cite{Peruzzo:2010p3875, Sansoni:2012p5030} and trapped atoms \cite{Karski:2009p2506, Genske2013p8085}. They exhibit different behaviour than classical random walks due to the interference of the quantum walker \cite{Nayak:2000p5322}. The quantum walk paradigm has been used to demonstrate that they are capable of universal quantum computing  \cite{Childs:2009p1332, Lovett:2010p2308}, including search algorithms \cite{Grover:1997p716} and more general quantum transport problems \cite{Kollar:2012p8332, Goswami:2010p2511}. The quantum walk comes in two varieties, the discrete-time (DTQW) and continuous-time (CTQW); the latter case will be the focus of this letter.

Recently an array of coupled waveguide channels with a down-conversion term was theoretically studied \cite{Solntsev:2012p5223} and experimentally demonstrated \cite{Kruse:2013p8361, Solntsev:2014p10020}. A classical pump drives a process creating down-converted light that then travels throughout the waveguide structure by evanescent coupling (the pump beam does not couple to other channels). This can be modelled by adding an extra term to the original CTQW Hamiltonian that converts two photons (walkers) from a single pump photon (the converse operation is also possible). During the \textit{driven} QW the walkers are created and annihilated and this in turn leads to very different dynamics when compared to the traditional \textit{passive} QW (PQW) which is restricted by a constant number of walkers. 

In this letter we investigate driven quantum walks. We connect the physical properties of the non-linear waveguide arrays with the traditional quantum walk formalism. Based upon the description of the system in the eigenmode basis, we show, that any driven QW can be decomposed into a PQW and an intricate input state. Furthermore, we are able to selectively pump spatial eigenmodes of the system, allowing for \textit{in situ} control of the QW properties. Finally, we take advantage of this unique property of driven QW to implement a search algorithm that demonstrates a quantum speed-up over a classical walker.

The CTQW is defined by a graph of coupled modes, such as a 2D lattice, and this graph topology can be encoded into a matrix $\mathbf{C}$ which describes the connections between the different modes of the system, as well as the on-site terms. The CTQW has the generic Hamiltonian,
\be
\hat{H}_C = \sum_{j,k} C_{j,k} \ad_j \aa^{}_k + \mbox{h.c.}, \label{ori_ham}
\vspace{-0.3cm}
\ee
where $\ad_j,\aa_j$ are the bosonic creation and annihilation operators respectively of the walker on the $j^{th}$ site of the graph and the evolution of the walk is given simply by the Schr\"odinger equation. Traditionally, the initial state is localized on a single mode e.g. $\left | \phi(t=0) \right \rangle = \ad_n|0\rangle$. The key to our subsequent analysis is to use the set of eigenmodes which diagonalize the original Hamiltonian~(\ref{ori_ham}), $\{\AA_k\}$. This will lead to (\ref{ori_ham}) being written in the form $\hat{H} = \sum_k \Omega^{}_k \AD_k \AA^{}_k$, where the $\{ \Omega_k \}$ are the eigenfrequencies, which in 1D lines and 2D lattices can be considered as a dispersion curve. The transformation from the original, physical basis $\{\hat{a}_k\}$ to the eigenbasis is given by the matrix $\mathbf{T}$, $\mathbf{A} =\mathbf{ T \, a}$. For the PQW the number of walkers in the eigenmodes of the system does not change. During the propagation, the phases between the eigenmodes change, leading to the well-known QW properties. 

We now add an extra term to this Hamiltonian that creates and destroys photons i.e. we change the number of walkers during the QW. This term will take one of two forms, 
\bea
\hat{H}_L = \sum_k \Gamma_{L,k}(t) \ad_k + \Gamma_{L,k}^*(t) \aa_k, \label{las_term} \\
\hat{H}_S = \sum_k  \Gamma_{S,k}(t) \ads_k + \Gamma_{S,k}^*(t) \aa^2_k. \label{squ_term}
\eea
We call these terms lasing and squeezing respectively,  as they are the Hamiltonians used for the generation of coherent states (created by a laser above threshold) and squeezed vacuum states \cite{Barnett_Radmore}. The lasing term can be realised in a DTQW with walkers added after each time step (to be studied later) and the second term was recently studied \cite{Solntsev:2012p5223, Kruse:2013p8361, Solntsev:2014p10020}. We only include one growth term at a time in our Hamiltonian and assume that these processes take place continually within the walk and are driven by an undepleted classical pump (i.e. a large amplitude coherent state) with a vacuum input state. The parameter $\mathbf{\Gamma}$ is the spatial pump shape and its time dependence will only depend on the pump frequency. The second term $\hat{H}_S$ is a down-conversion process of two photons from a single pump photon. In the main text of the letter we focus on the lasing QW. Similar results for the squeezing QW may be found in the supplementary information \cite{dqw_supp_info}.

\begin{figure}
\includegraphics[width=.45\textwidth]{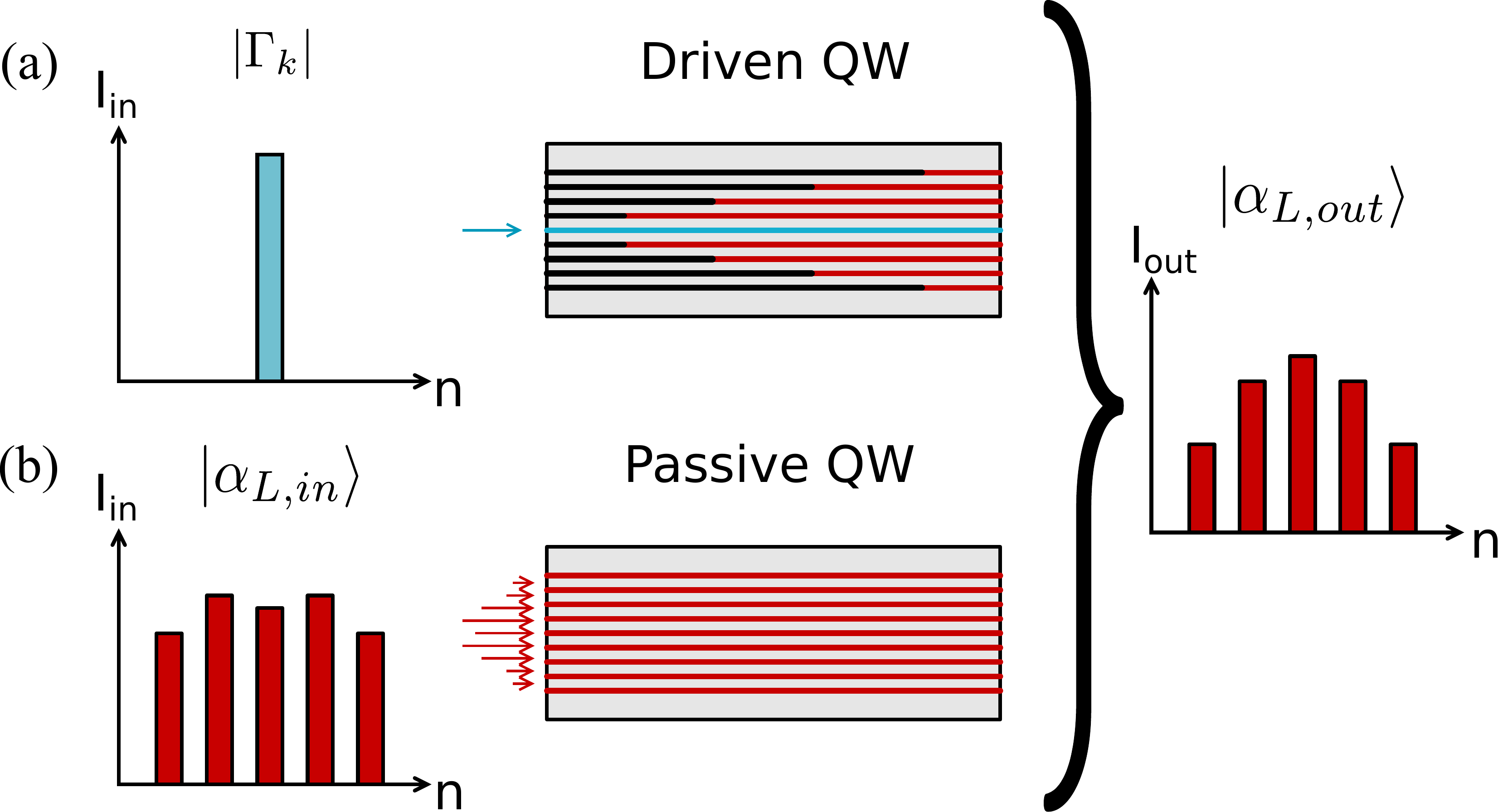}
\caption{Any driven QW (a) can be decomposed into an intricate multimode state followed by a passive QW (b). The output states of the two system will be the same. }
\label{fig:QW-interpretation}
\end{figure}

Using the transformation $\mathbf{T}$ the complete Hamiltonian $\hat{H}_C+\hat{H}_L$ in the eigenbasis is,
\be
\hat{H}=  \sum_k \Omega_k \AD_k \AA_k + \sum_{k} S_{k}(t) \AD_k + \mbox{h.c.}
\label{eq:eigenmode-Hamiltonian}
\ee
where $\mathbf{ S_L = T^{-1} \Gamma_L }$ ($\mathbf{ S_S = T^{-1} \Gamma_S T }$ for the squeezing term as $\Gamma_S$ is a matrix ). We now move to the interaction picture of the dynamics using the transformation \cite{Barnett_Radmore} $\hat{U} = \prod_k \exp \left(i \Omega_k \AD_k \AA_k t \right) $ which allows us to rewrite equation \eqref{eq:eigenmode-Hamiltonian} as
\be
\hat{H}_{int} = \sum_{k} S_{k} (t)\AD_k e^{i \Omega_k t } + \mbox{h.c.}
\ee
Integrating over time, $ \mathbf{ z} = \int dt' \mathbf{ S }(t') $, and using \cite{Ma:1990p3591} to disentangle the Schr\"odinger evolution operator, when we convert back to the original operator basis, $\{ \aa_k \}$ this yields an output state of the form,
\be
\begin{aligned}
\ket{\alpha_{L,out}}&= \underbrace{\exp \left ( -i t \sum_{k,k'} C_{k,k'} \ad_k \aa^{}_{k'} \right)}_{\hat{U}_{PQW}} \underbrace{\exp \left ( -i \sum_{k} z'_{k}  \ad_k \right )\ket{0}}_{\ket{\alpha_{L,in}}}\\
						&= \hat{U}_{PQW} \ket{\alpha_{L,in}}\, ,
\end{aligned} 
\label{evo_op}
\ee
where $ \hat{U}_{PQW}$ is the evolution operator of the PQW, $\ket{\alpha_{L,in}}$ is the `initial state' and $z'_k$ is the expression for $\mathbf{z}$ in the physical basis. This result is illustrated in figure \ref{fig:QW-interpretation}. We can decompose any driven QW in figure \ref{fig:QW-interpretation}(a) into a multimode coherent state that is then launched into the original CTQW with Hamiltonian evolution $\hat{U}_{PQW}$, as shown in figure \ref{fig:QW-interpretation}(b), leading to the same output state. A more detailed derivation may be found in the supplementary material \cite{dqw_supp_info}.

We can apply a similar theory to the other growth term  $\hat{H}_S$ and arrive at an identically structured evolution of the walk i.e. first we create a multimode squeezed state then we evolve this state through the quantum walk. Note that the state creation and walk are not independent; both are a function of the overall time/length of the walk. We can lift this restriction on the input states to a certain extent by starting with a non-zero state at the beginning of the walk. If we had started with a non-vacuum initial state then additional terms would have to be taken into account, changing equation (\ref{evo_op}) but retaining the same structure of state creation followed by quantum walk.

The `initial' state in (\ref{evo_op}) is determined by two factors, both contained within the matrix $\mathbf{S}$. Firstly, the pump shape which is simply the absolute value of the elements of $\mathbf{S}$, $|S_{k}|$, (or $|S_{k,k'}|$ for $\hat{H}_S$)  and determines which eigenmodes have a non-zero growth. The second factor is the phase-matching that occurs due to the time-dependence of $\mathbf{S}$, typically of the form $e^{-i\omega_p t}$, where $\omega_p$ is the pump frequency. More complicated time dependencies would not alter the interpretation presented here. When we insert the pump shape into the interaction Hamiltonian,
\be
\hat{H}_{int}=\sum_k S_k \exp\left(i(\Omega_k-\omega_p)t\right)\hat{A}^\dag_k + \mbox{h.c.},
\ee
it becomes clear that we have to fulfil the eigenmode phase-matching condition $\omega_p = \Omega_k$ for $\hat{H}_L$ (or $\omega_p = \Omega_k + \Omega_{k'}$ for $\hat{H}_S$). When integrating over time phase-matched eigenmodes grow linearly in time whereas non-phase matched modes oscillate depending upon the size of the phase mismatch. These two factors give some control over which eigenmodes are created during the QW. 

The main difference between the driven quantum walk described here and the original CTQW can be easily seen in the nature of the eigenmodes. In the CTQW the amplitudes of the eigenmodes are fixed at the start of the walk and only the phases change in time. In our driven QW we can choose the eigenmode(s) we want to create by changing the pump's spatial shape and frequency to drive and phase-match combinations of eigenmodes, thus the amplitudes (and phases) of the eigenmodes change. For longer walks phase-matching is the more significant factor, where the majority of walkers are created in eigenmodes which are phase-matched (or almost phase-matched). 

As an example we look at the evolution and dynamics of the driven quantum walk in a 1D array of $N$ coupled oscillator modes,  
\be
\hat{H} = \omega \sum^N_{k=1} \ad_k \aa^{}_k  +  C \sum^{N-1}_{k=1} \ad_k \aa^{}_{k+1} + \mbox{h.c.}.
\ee
a topology which has been studied previously in CTQW. The eigenmodes $\{\hat{A}_j\}$ in the finite-system case are given by $\hat{A}_j=\sqrt{2/N}\sum_k \sin(j k / (N+1)) \hat{a}_k$ and eigenfrequencies $\Omega_j=\omega+2 C \cos(j \pi /(N+1))$. Here we use the lasing growth term (\ref{las_term}), with a waveguide array consisting of 51 modes ($-25 \le k \le 25$), (the values we use are $\omega=1$, $C=0.5$, $\Gamma_0 = 1$ and we run for a time of $t=20$ in dimensionless units) and we only pump a single physical mode $k=0$, the middle mode in the chain, with the pump frequency $\omega_p=\Omega_1$.
\begin{figure}
\begin{center}
\includegraphics[trim= 10cm 11.5cm 10cm 12.5cm , scale=0.7]{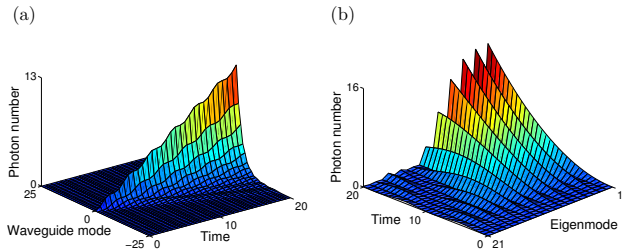}
\caption{(a) Photon number in the physical basis and (b) eigenbasis during a lasing driven QW. The total number of photons is equal in both bases. }\label{fig_qwlas}
\end{center}
\end{figure}

Figure \ref{fig_qwlas} shows the average photon number dynamics during the walk in the physical basis (2a) and the eigenmode basis (2b). The nature of phase-matching can be seen in Fig \ref{fig_qwlas}b as the phase-matched mode ($j=1$) will continue to grow indefinitely, while the non-phase-matched modes ($j \ne 1$) grow in number then decrease (and will continue to oscillate). As the walk continues the phase-matched modes will drown-out the other modes. The total photon number grows quadratically in the lasing case and exponentially in the squeezing case for the phase-matched modes.

\begin{figure}
\begin{center}
\scalebox{0.25}{\includegraphics{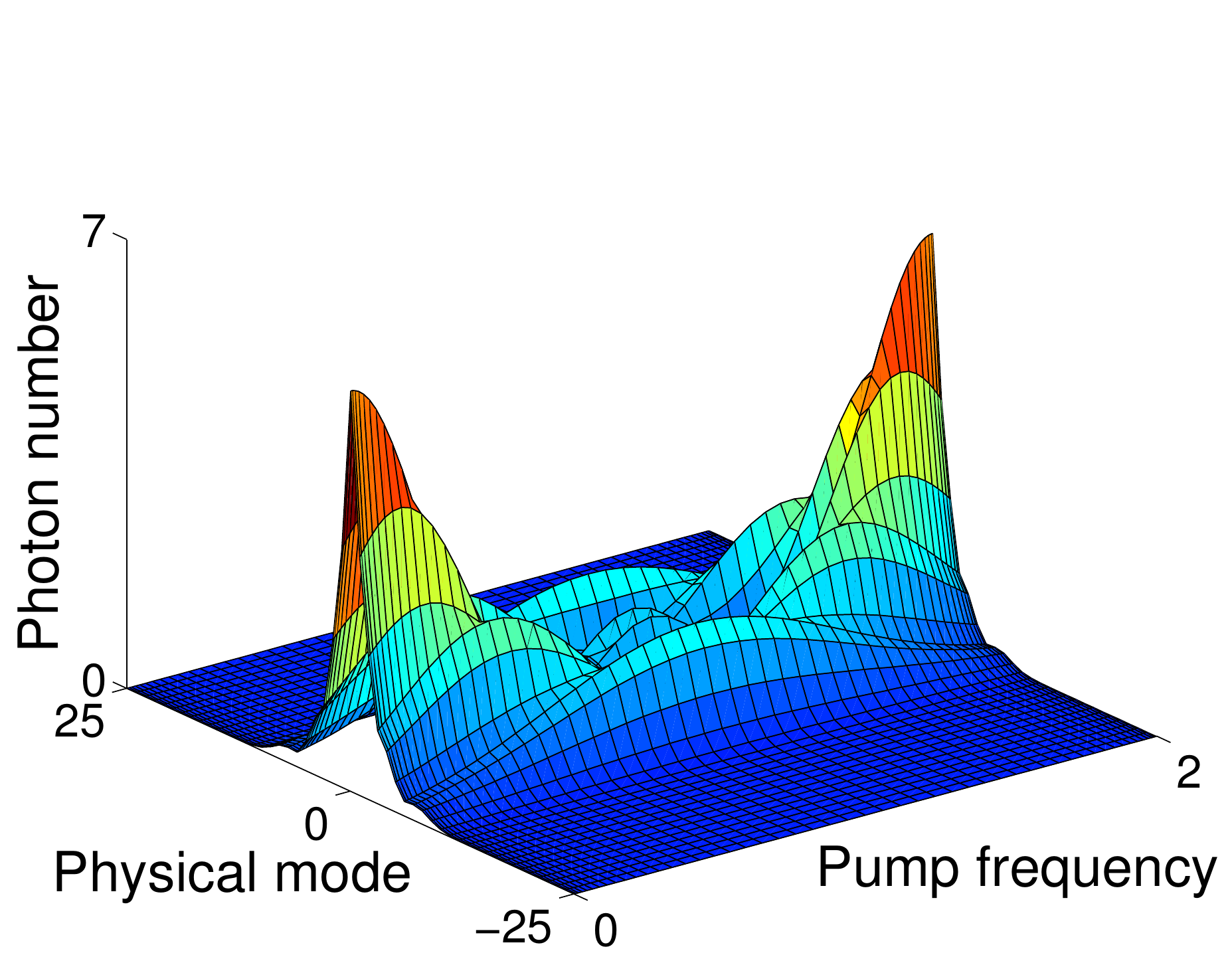}}
\end{center}
\caption{Photon number in the physical basis at the end of the lasing QW in each waveguide vs. pump frequency. Pump frequency is in the range $[0,2]$.}\label{fig_qwlas2}
\end{figure}

The distribution of the walkers' position will depend upon the pump frequency, shown in figure \ref{fig_qwlas2}. The output distribution at the end of the walk changes as we change the pump frequency, while keeping the walk length constant, over the range of eigenfrequencies. We can see that the final output distribution changes from a peaked structure to one that spreads out as the frequency changes. We can see from the plots above that the dynamics of the driven walks do not resemble those of the traditional quantum walk. The walkers tend to stay localised about the channel that the pump beam is present in, especially in the case where squeezing is present (shown in the supp. info.). Using our interpretation of the driven QW, equation ~(\ref{evo_op}), our initial state will generally be extended over several waveguide channels. As shown in \cite{deValcarcel:2010p3858}, this leads to very different dynamics when compared to an initial state localised in a single mode, which usually has two lobes traveling away from the input site at speeds $\pm \sqrt2 C$. Similar results for the squeezing growth term are included in the supplementary material \cite{dqw_supp_info}. 

A typical measure of quantum walk dynamics is the variance of the walkers position distribution as the walk evolves in time ($\sigma^2(t) = \sum_x x^2\bar{n}_x(t)$) as this is markedly different from classical diffusion. In these type of walks here we have two sources of growth: from the actual spreading of the walkers position and from the change in mean walker number ($\bar{n}_x(t)$). In the example shown here the variance grows as $ t^3$, though when we re-scale this variance by the average photon number (so we always have what can be considered the position-distribution of a single photon) which grows as $t^2$, we instead see linear growth of the spatial-only variance. This can be explained due to the walkers being created in a single mode. Different pump frequencies lead to similar regimes of variance-growth.

One of the main applications of quantum walks are search algorithms, such as the Grover search \cite{Grover:1997p716, Shenvi:2003p5446}. Here a particular initial quantum state evolving under the DTQW can find a marked vertex in a time $ t \propto \sqrt{N}$, where $N$ is the number of vertices. This represents a speed-up over classical search algorithms. There are other QW search algorithms that are closely related \cite{Childs:2004p5474, Farhi:1998p5411, Feldman:2010p2811, Cottrell:2014p9124}. In a traditional QW search algorithm the walker starts in a spatially extended state over all vertices and the dynamics cause it to oscillate in-and-out of the target vertex, as illustrated in figure \ref{dqw_search_dyn}(a), thus only being found there at certain times. Here we present an alternative scheme based on the driven QW. The strengths of this scheme is that we start from vacuum, thus no complicated initial state, and we continually pump walkers into the walk so the walker-oscillation never occurs, thus the walker can be measured at any time (after a minimum time to drown out other eigenmodes).

For our scheme, we consider a topology  where there is a `pump' or `entrance' mode, $\aa_p$, and a marked `defect' or `exit' vertex, $\aa_d$. Our aim is to find the defect vertex by having a larger number of the walkers there than the other vertices. The way we achieve this is by matching the pump frequency, $\omega_p$, to that of an eigenmode, $\Omega_D$, that is predominantly a combination of the defect vertex and the pump vertex e.g. $\AA_D \approx \mu_{D,p} \aa_p + \mu_{D,d}\aa_d $ (where $\mu_{D,.}$ is the weight of the physical modes in the composition of the eigenmode) and have a very low weight of other modes i.e. $\mu_{D,p} =\mu_{D,d} =1/\sqrt{2}$. (These conditions are closely related to centro-symmetric matrices \cite{Cantoni:1976p8103} which have been shown to help efficient quantum transport \cite{Walschaers:2013p8632}. )

With this, our scheme depends upon 3 factors. First is the shape of the eigenmode $\AA_D$, which must comprise a large proportion of the entrance and exit modes compared to other system modes. Next, will be how these proportions change with increasing system size. Finally there is the distance between two eigenfrequencies, determining the minimum  phase-mismatch between eigenmodes, which gives the minimum time one would have to wait before measuring the system in order to drown-out the other non-phase-matched modes. 

\begin{figure}
\centering
\includegraphics[trim=22cm 9cm 15cm 15cm, scale=0.6]{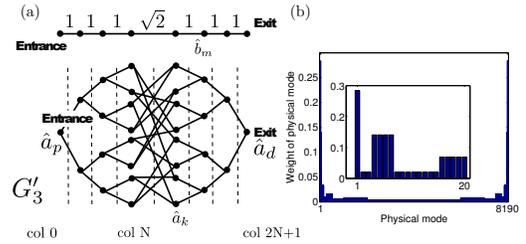}
\caption{ (a) GTG of depth $N=3$, with columns emphasised by dashed lines. (b) Shape of eigenmode ($\AA_j$) we wish to phase-match. Inset is zoomed on first 20 modes. }\label{fig_file_gtg_eigen}
\end{figure}

\begin{figure}
\centering
\includegraphics[trim=4.5cm 11.5cm 0cm 11.5cm, scale=0.7]{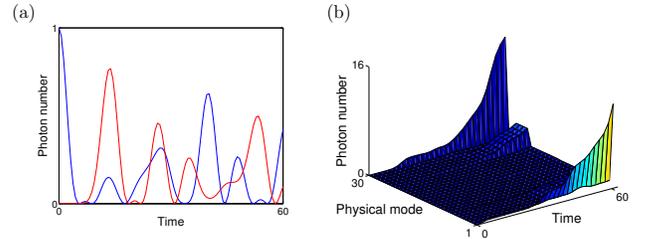}
\caption{(a) CTQW of a single photon on the GTG with a photon starting on the entrance mode, with the oscillations clearly seen. For clarity only the entrance/exit modes are shown (blue is entrance, red is exit).  (b) Driven search on the same graph with phase-matching to a single eigenmode. }\label{dqw_search_dyn}
\end{figure}

We illustrate this scheme with the glued-trees graph (GTG) \cite{Childs:2003:EAS:780542.780552}, which is characterised by its depth $N$ (total number of vertices $2^{N+2}-2$) whose topology of vertices, $\{\hat{a}_k\}$, can be seen in figure~\ref{fig_file_gtg_eigen}a for $N=3$ (eigenmodes are $\{\hat{A}_k\}$). It has been shown for this particular graph that a classical walker takes an exponential time to move from the entrance vertex $\hat{a}_p$ to the exit vertex $\hat{a}_d$ but a quantum walker takes a polynomial time. Here we provide some analytical and numerical evidence that our method can also traverse in a time that scales polynomially. More details can be found in supplementary material \cite{dqw_supp_info}.

To analyse this system we use the column representation (the columns are highlighted by the dashed lines in fig.~\ref{fig_file_gtg_eigen}a) which maps the full set of vertices, $\{\aa_k\}$, to a linear chain of $2N+2$ oscillator modes $\{\hat{b}_m\}$ (eigenmodes $\{\hat{B}_m\}$), as sketched in figure \ref{fig_file_gtg_eigen}a. The coupling coefficient between modes is $C_{m,m+1}=1$  for all but the central two for which it is $C_{N,N+1}=\sqrt{2}$. The linear chain modes, $\hat{b}_m \propto \sum_{k, \hat{a}_k \in \mathrm{col~m}} \hat{a}_k$ represent $2^{m}$ modes in column $m$ of the full graph for $m=0,..,N$ and is mirrored for $m=N+1,..,2N+1$. At the ends of the graph $\hat{b}_0 = \hat{a}_1$ and $\hat{b}_{2N+1} = \hat{a}_{2^{N+2}-2}$. We will examine eigenvectors $\hat{B}_m$ that have a large weight at the ends. 

Expressions for the eigenvectors of the linear chain, $\hat{B}_m$, can be written down analytically \cite{Ide:2014p9615} and are evaluated at the eigenvalues of the system. These analytical expressions are identical for the 1D chain with/without the central-coupling difference, although the eigenvalues are different for each system. The eigenvectors for the 1D chain were stated above. By inspection, the best performing eigenvector can be approximated by $\hat{B}_N = 1/\sqrt{N} \sum_m \sin( N \pi m / (2N+1)) \hat{b}_m \approx 1/\sqrt{N} \sum_m \sin(\pi m / 2 ) \hat{b}_m$. Due to each column representing many GTG vertex modes, we can renormalise $\hat{b}_m$ according to the number of $\hat{a}_k$ it represents (e.g. $2^{m}$ for $m=0,..,N$). This gives a large weight at the ends of the linear chain, or at the entrance and exit of the graph, and this behaviour can be seen in figure~4 in the supp.~info. Also, the weight scales as $1/\sqrt{N}$ which fulfils two of our criteria. In the supp.~info. \cite{dqw_supp_info} we also show additional numerical evidence that the weights of the entrance/exit modes decrease as $1/\sqrt{N}$ for the full graph for depth up to N=11, (total of 8190 vertices). Figure \ref{fig_file_gtg_eigen}b shows the shape of the eigenmode that we wish to create, with large weights on both the entrance and exit. 

Finally, the eigenfrequencies of the system are approximately satisfied by $\Omega_j=\cos(j\pi/(2N+1))$, so the phase mismatch between $\omega_p = \Omega_N$ and $\Omega_{N\pm1}$ decreases polynomially, which only increases the time we have to wait before measuring polynomially. The eigenmode with the smallest phase mismatch $\Delta$ between the eigenfrequency and pump frequency oscillates with a period $1/\Delta$ thus gives us the time we have to wait for this eigenmode be drowned out relative to the phase matched mode.

Figure \ref{dqw_search_dyn} is a direct comparison between the traditional QW (a), which shows the oscillation of the photon between the entrance and exit modes, and the driven QW (b), which shows the continuous growth of photons in both the entrance and exit modes.

In conclusion we have introduced and discussed a new type of quantum walk, one where the walkers are created and destroyed coherently during the walk. We have placed the walk dynamics into a version more easily interpreted in the traditional sense of quantum walks, which is state creation followed by the quantum walk. This leads to a quantum walk of multimode coherent states or squeezed states depending upon the type of source Hamiltonian term used. An experimental realisation of this has already been carried out \cite{Kruse:2013p8361, Solntsev:2014p10020}. This type of walk is very different due to the active process of choosing which modes are being created, which can lead to interesting and novel transport properties in such systems. Next, we have suggested a search protocol that this process could perform. Here we phase-match a special eigenmode of a disordered system in order to discover where the marked vertex is. This leads to a growth of walkers in the defect mode which allows it to be identified. This has the advantages that there is no need for a complicated initial state and that the walkers do not oscillate away from the defect mode, as in the traditional quantum search. We will study other aspects of these walks, such as there performance with disorder, in future work.

\begin{acknowledgments}
CSH and IJ received financial support from grants RVO 68407700 and GA\v{C}R 13-33906S.
\end{acknowledgments}

\onecolumngrid
\section*{Supplementary Information}

\section{Derivation of the decomposition}\label{app_derivation}

Here we derive the decomposition of the driven quantum walk into an initial state and passive quantum walk. Our method to compare the two walks relies on solving the Hamiltonian dynamics of the walk and re-arranging the evolution operators, which we illustrate with the squeezing term but the same analysis leads to an identical result for the lasing term. We first diagonalise the linear part of the Hamiltonian, 
\be
\hat{H} = \sum_{j,k} C_{j,k} \ad_j \aa_k \rightarrow \sum_k \Omega_k \AD_k \AA^{}_k
\ee
and use the transformation $T$ that relates the two sets of operators to transform the growth terms. The final Hamiltonian in the eigenbasis is,
\be
\hat{H}_A =  \sum_k \Omega_k \AD_k \AA_k + \sum_{k,k'} S_{k,k'}(t) \AD_k\AD_{k'} + \mbox{h.c.}
\ee
where $\mathbf{S}(t) = \mathbf{T^{-1} \Gamma_s}(t) \mathbf{T}  $. We now move to the interaction picture of the dynamics using the transformation \cite{Barnett_Radmore},
\be
\hat{U} = \prod_k \exp \left(i \Omega_k \AD_k \AA_k t \right) 
\ee
which gives,
\be
\hat{H}_{int}(t) = \sum_{k,k'} S_{k,k'}(t) \AD_k\AD_{k'} e^{i (\Omega_k + \Omega_{k'}) t } + \mbox{h.c.}
\ee
The state evolution in the interaction picture is given by,
\be
| \phi \rangle_{int} = \exp \left (-i \int_0^t dt' \hat{H}_{int}(t') \right ) | 0 \rangle_{int} 
\ee

When we move back to the Schr\"odinger picture, 
\bea
|\phi \rangle &=& U^\dag | \phi \rangle_{int} = U^\dag \exp \left (-i \int_0^t dt' \hat{H}(t') \right ) U | 0 \rangle \nn \\ 
&=& U^\dag \exp \left (-i \int_0^t dt' \hat{H}_{\mbox{int}}(t') \right)| 0 \rangle
\eea
and we integrate over $t$. The first term is of the form
\be
\exp \left (-i \int_0^t dt' \hat{H}_{\mbox{int}}(t') \right) = \exp \left ( -\frac{i}{2}\sum_{k,k'} z_{k,k'} \AD_k \AD_{k'} + \mbox{h.c.} \right )  = \exp \left ( \frac{(\AD)^T \mathbf{z} \AD}{2} - \frac{(\AA)^T \mathbf{z^*} \AA}{2}   \right ) 
\ee
where $\mathbf{z}$ is a $N \times N$ matrix and the operators here  are considered as row and column vectors of the individual modes and are used to write the operator in a compact way. Now, using \cite{Ma:1990p3591}, we will disentangle this operator and put it in the form,
\be 
\exp \left ( \frac{(\AD)^T \mathbf{z} \AD}{2} - \frac{(\AA)^T \mathbf{z^*} \AA}{2}   \right )  = \exp{\frac{(\AD)^T \mathbf{z' }\AD}{2}  } \,\exp{f(\mathbf{z})\AD \AA} \,\exp{ \frac{(\AA)^T \mathbf{z'^*} \AA}{2} }\nn
\ee
where $f\mathbf(z)$ is an unimportant function as the first two terms will act on the vacuum and can therefore be neglected.

The evolution of the state can now be written as,
\be
|\phi(t) \rangle  = \exp \left ( -i t \sum_k \Omega_k \AD_k \AA^{}_k \right ) \exp \left ( -i \sum_{k,k'} z'_{k,k'}  \AD_k \AD_{k'} \right ) | 0 \rangle
\ee

If we convert back to the original operator basis, $\{ \aa_k \}$, it becomes, 
\be
|\phi(t) \rangle  = \exp \left ( -i t \sum_{k,k'} A_{k,k'} \ad_k \aa^{}_{k'} \right) \exp \left ( -i \sum_{k,k'} z''_{k,k'}  \ad_k \ad_{k'} \right ) | 0 \rangle
\ee
Looking at this we can interpret this process as creating a multimode squeezed state that is then launched into the original CTQW with Hamiltonian evolution.

\section{Squeezing term}\label{app_sque_term}

Here we use the two-photon downconversion term,
\be
\hat{H}'_S = \Gamma_0 e^{-i\omega_p t}  \hat{a}_0^{\dag 2} + \mbox{h.c.}
\ee
and keep the same 1D nearest-neighbour Hamiltonian as before ($\Gamma_0=0.1$). This system can be experimentally realised by Kruse {\it et. al.} \cite{Kruse:2013p8361} if narrow-band filters had been placed at the end of the waveguide array. 

\begin{figure}[phtb]
\begin{center}
\scalebox{0.25}{\includegraphics{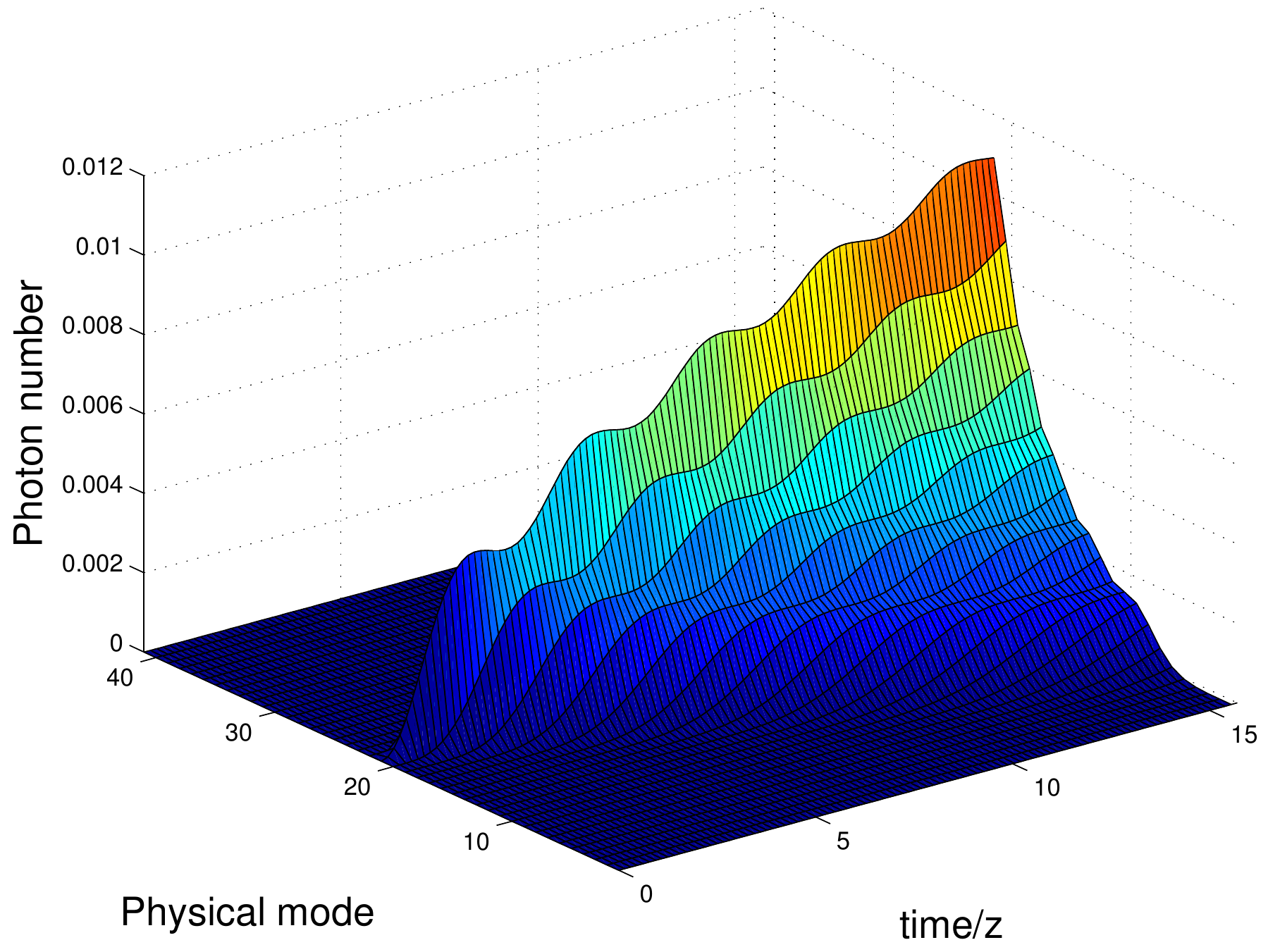}}
\end{center}
\caption{Average photon number in the physical basis during a squeezing quantum walk. The pump frequency is the middle eigenfrequency.  }\label{fig_qwsqu}
\end{figure}

\begin{figure}[phtb]
\begin{center}
\scalebox{0.25}{\includegraphics{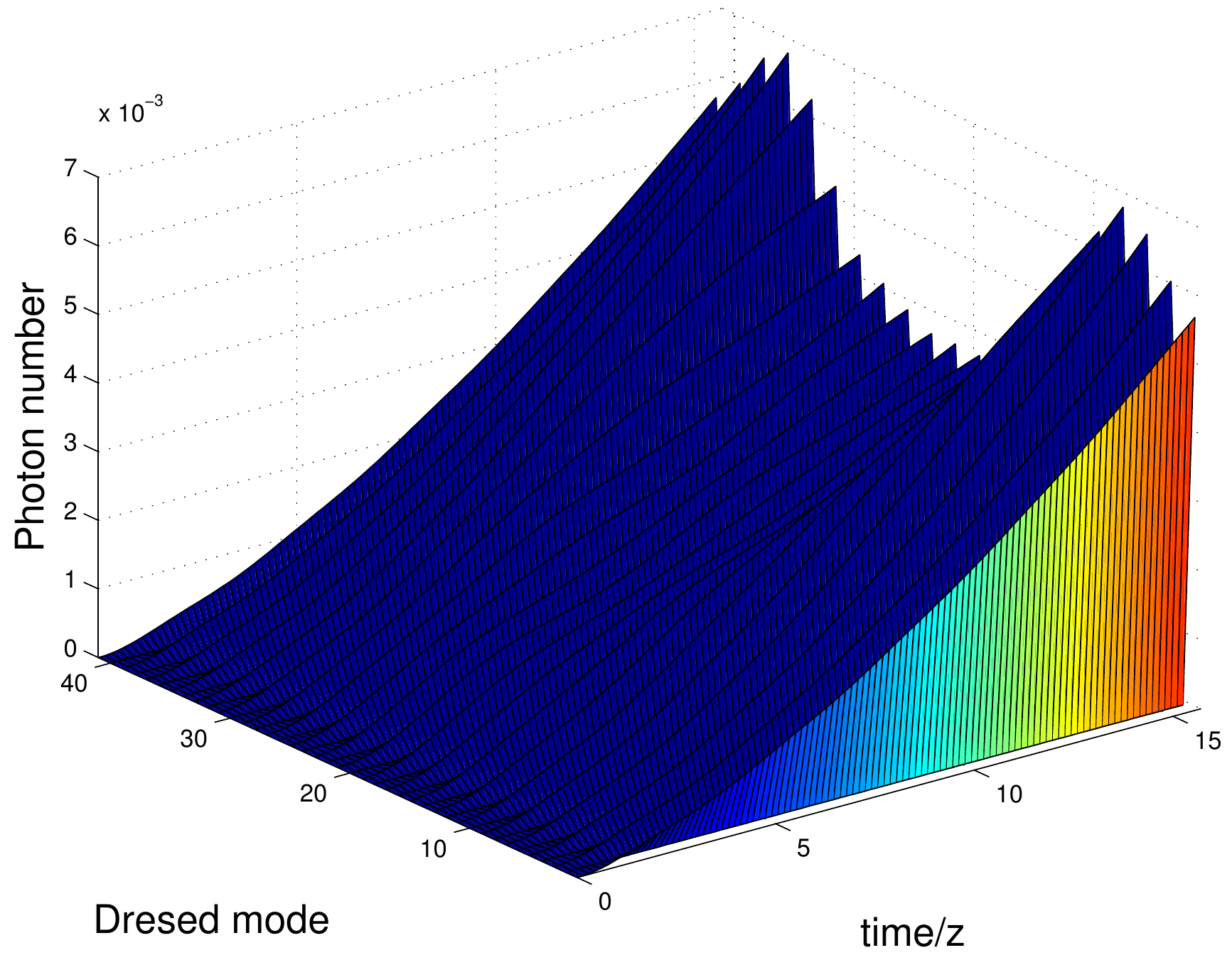}}
\end{center}
\caption{Average photon number in the eigen basis during a squeezing quantum walk. The pump frequency is the middle eigenfrequency. }\label{fig_qwsqu_drm}
\end{figure}

\begin{figure}[phtb]
\begin{center}
\scalebox{0.25}{\includegraphics{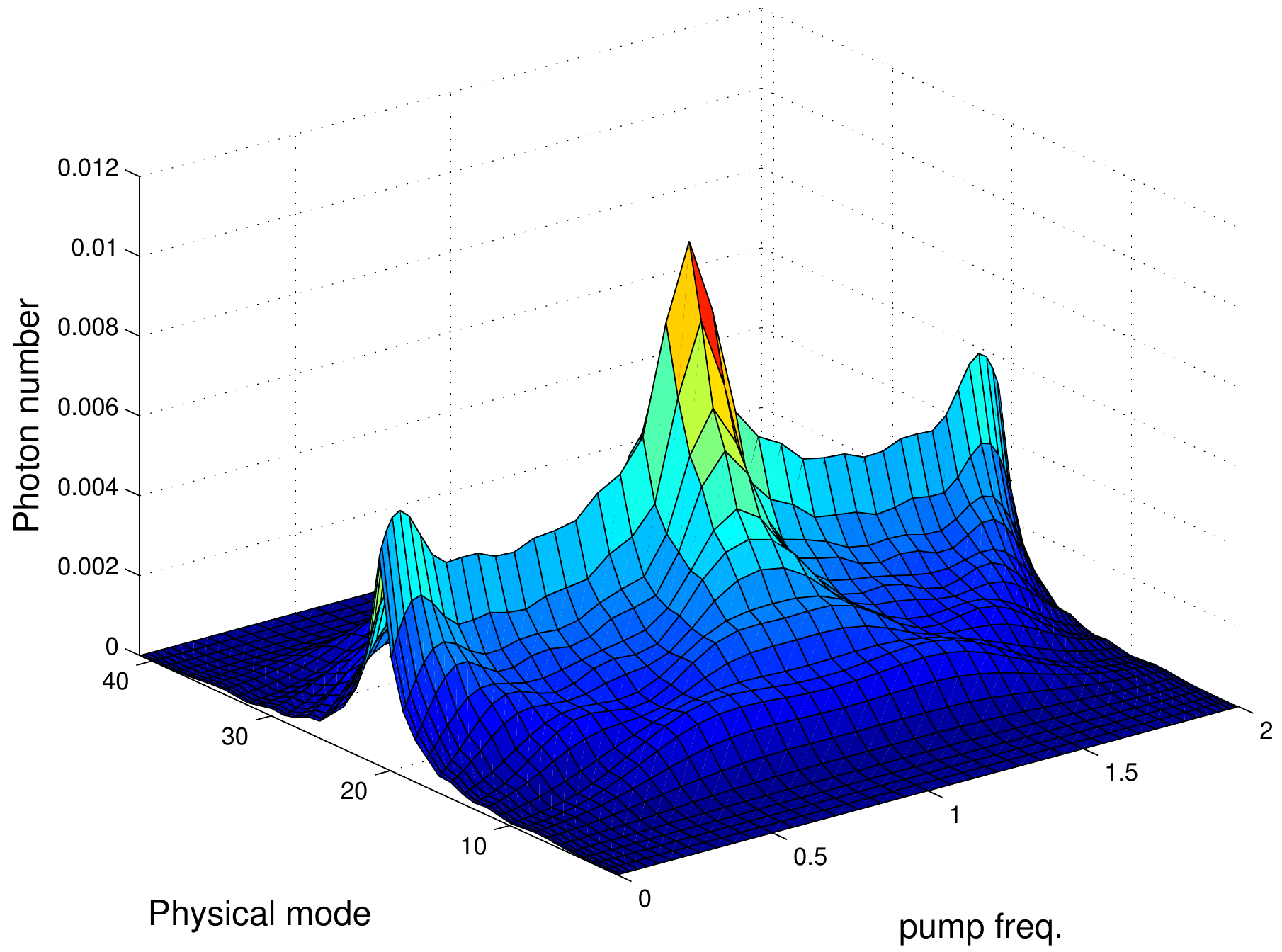}}
\end{center}
\caption{Average photon number in the physical basis at the end of a squeezing quantum walk as the pump frequency is varied. Pump frequency in this case is double the eigenfrequencies of the system. }\label{fig_qwsqu2}
\end{figure}

Below we plot the evolution of typical walk dynamics. These plots mirror those examined in the previous section. Figure~\ref{fig_qwsqu} shows the evolution of the photon number during a walk and this figure can be compared to figs. 7 and 8 in \cite{Kruse:2013p8361}, which show the output at the end of the quantum walk. Figure~\ref{fig_qwsqu_drm} shows the same walk in the eigenmode basis.   Figure~\ref{fig_qwsqu2} shows how the output distribution changes depending upon the pump frequency and it can be seen that the distribution is always localized in the channel that is pumped. 

\section{Glued trees graph}\label{app_gtg}

In this section we describe the eigenvalues and eigenvectors of the glued-trees graph, whose mode topology shown in figure~4a of the main text. The system can be divided into $2N$ columns and the total number of modes is $2^{N+2}-2$. As described in \cite{Childs:2003:EAS:780542.780552} this system can be mapped to a 1D linear chain, where each mode in the chain represents all the modes of each individual column of the original graph. Thus the linear chain has 2N modes with constant coupling $C$ between each mode, apart from a defect in between the middle two modes N,N+1 which is $\sqrt{2}C$.

Ref. \cite{Ide:2014p9615} gives us a way to analytically calculate the eigenvectors and eigenvalues of the 1D chain. The eigenvalues, $\lambda_j$, satisfy the equation, 
\be
U_N(\lambda_j) = U_{N-1}(\lambda_j),
\ee
where $U_N$ is the Chebyshev polynomial of the second kind of degree $N$ (compared to $\lambda_j = 2C\cos(j\pi/2N+1)$ for the linear chain without the defect). The eigenvectors are given by functions whose form is laid out in \cite{Ide:2013p8901}, which is the same for both linear chains with and without the defect. The difference is that the functions are evaluated at the eigenvalues of the two systems, which are different. As the size of the system increases, $N\rightarrow \infty$, the eigenvalues of each system tend to each other, thus the eigenvectors will tend to one another also. 

\begin{figure}[phtb]
\begin{center}
\includegraphics[trim= 11cm 12cm 11cm 11.5cm , scale=1]{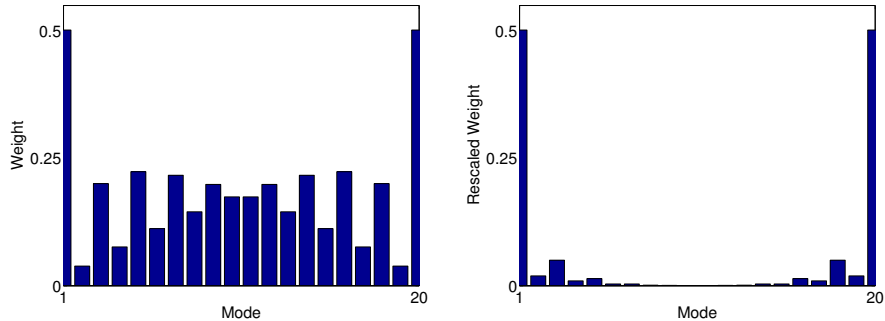}
\end{center}
\caption{Shape of an eigenmode of the linear chain (left) and the same eigenmode that takes into account the number of modes in each column (right).}\label{fig_eigenmodes_chain}
\end{figure}

Looking the eigenvector of the 1D chain, $v_{j,k}=\sqrt{2/2N}\sin(j k \pi/(2N+1))$, that has a large weight at the ends of the chains $k=N$ for large $N$, $v_{j,N}=\sqrt{1/N}\sin(j \pi/2)$. In figure \ref{fig_eigenmodes_chain} we plot the eigenvector of the 1D chain (with central defect) with the largest weights at the end and then that eigenvector re-scaled by the number of mode in that channel. It is clear from the figures that eigenvectors exist that are predominantly located on the entrance and exit modes (i.e. the ends of the linear chain). These plots should be compared with the right side of fig~\ref{fig_file_gtg_eigen} that shows the eigenmode of the full graph of 8190 modes. 

\begin{figure}[phtb]
\centering
\includegraphics[trim=3.5cm 11.5cm 2cm 12cm, scale=0.6]{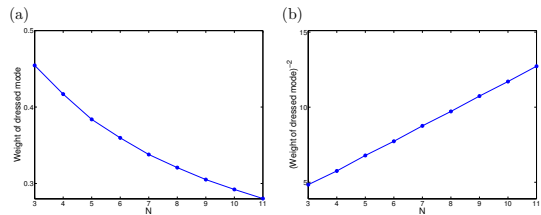}
\caption{(a) Numerically calculated  weight of eigenmode $\AA_D$ and (b) (weight)$^{-2}$ vs. depth of full GTG.}\label{weight_fig} 
\end{figure}

In figure~\ref{weight_fig} we plot how the weight of the entrance/exit mode in the composition of $\AA_D$ scales with increasing system size for the full GTG up to depth N=11(total of 8190 vertices). This numerical analysis of this data shows that is does indeed have a $1/\sqrt{N}$ dependence.

\bibliography{draft_prl.bib}

\end{document}